\documentstyle[epsfig,12pt]{article}     
\begin{document}
 \begin{center}
{\bf A new view of the $0^-$ and $2^+$ glueballs}
\vskip 1cm

    { D.V. Bugg and B.S. Zou}

    {Queen Mary and Westfield College, London E1 4NS, UK}

\vskip 1cm
    {\bf Abstract}

\end{center} 
\vskip 1cm

\baselineskip=24pt
Data on $J/\Psi \to \gamma (\rho \rho)$ demand a broad $0^-$ $\rho \rho $ 
signal, attributable to a glueball  with a mass of 1750--2100 MeV.
Decays of this broad state to $\rho\rho$, 
$\omega \omega$, $K^*\bar K^*$ and $\phi \phi$
channels agree well with flavour blindness.
The narrow $\iota(1440)$ may be attributed to mixing between the glueball
and the $s\bar s$ radial excitation. 
The latter is pushed down in mass by
repulsion between the two levels. We conjecture that the $2^+$ glueball may
likewise be broad. Mixing between it and the $2^3P_2~q\bar q$ and $s\bar s$ 
radial excitation can explain the appearance of $f_2(1565)$ and a $2^+$ 
$\Theta (1710)$ at masses lower than anticipated. 
Mixing with higher $q\bar q$ states can
explain the $2^+$ resonance at 1920 MeV and also $\phi \phi$ signals 
observed by Etkin et al. and by the JETSET collaboration.

\vskip 6cm
We begin by correcting a conceptual flaw in some of our earlier work
and that of others.
It has been common practice to fit data on the $\iota (1440)$ using a
Breit-Wigner amplitude with a constant width. 
For a narrow resonance, this is a
good approximation. However, in $J/\Psi \to \gamma (\pi ^+ \pi ^- \pi ^+
\pi ^-) $ a broad signal is seen extending from the $\rho \rho$ threshold
to 2.3 GeV or more. 
In our earlier work [1], we followed the practice of using a Breit-Wigner 
amplitude with a constant width. This
amplitude succeeded in fitting the data with a $0^-$ resonance having a mass of
1420 MeV, alluring close to the $\iota (1440)$; but it had a width of 160 MeV,
much larger than the average value of 60 MeV quoted by the Particle Data 
Group [2]. 
The broad $0^-$ signal in $\rho \rho$ arose from a 
$k^3$ P-state phase space factor and a modest form factor fitted 
to the momentum dependence of the data. 
 Here $k$ is the momentum
of each $\rho$ in the $4\pi$ rest system.  

However, for a broad state, a Breit-Wigner amplitude of constant width is a
bad approximation. It is necessary instead to use a Flatt\'e form [3] for the
amplitude $f$ of the resonance. 
For decay channel $i$, after integrating over the phase space for decay,
\begin {equation}
f_i(s)  = \frac {B_J(s)B_i(k)\sqrt {\Gamma _i}}{M^2 - s - iM(\sum_i \Gamma _i)}.
\end {equation}
Here $M$ is the mass of the resonance, $s$ the invariant mass squared 
of observed
particles, and $\Gamma_i$ are the decay widths to each channel; $B_J$ and $B_i$
are form factors, which will be given later. 

We have returned to an analysis of the data on $J/\Psi \to \gamma (4\pi )$
using self-consistent $\Gamma _i(s)$ in the Flatt\'e formula. We reach the
unambiguous conclusion that these data cannot be fitted consistently with the
narrow $\iota (1440)$ observed in $\eta \pi \pi$ and $K\bar K \pi$ channels
using just a single resonance. 
This is illustrated in Fig. 1, which makes a comparison with Mark III data on
$K\bar K \pi$ [4] and $4\pi$ [5] channels. 
If the resonance is made wide enough to fit
the $4\pi$ data, it is far too wide to fit $K\bar K \pi
$ (or $\eta \pi \pi$); this is
illustrated by the full curves. 
Conversely, if it is made narrow enough to
fit the $K\bar K \pi$ data, it gives a narrow peak at 1430 MeV in the $4\pi$
channel (shown by the dashed curves), despite the rapidly increasing $4\pi $
phase space.
The only way of fitting the data successfully, as described below, is to 
combine a narrow resonance at $1430 \pm 15$ MeV in 
$\eta \pi \pi$ and $K\bar K \pi$ 
with a broad signal in $\rho \rho$ (and other channels). 

Both signals appear strongly in $J/\Psi$ radiative decays. It is natural to
assume a connection with a $0^-$ glueball. Such a state is predicted by Lattice
QCD calculations at almost the same mass as the $2^+$ glueball [6]. 
We propose that it is to be identified with the broad $\rho \rho $ signal. 

In order to place the mass of the $\iota (1440)$ in context, we consider 
masses in the $0^-$ radially excited nonet. 
There have been several reports of a narrow $\eta (1295)$ decaying to $\eta \pi
\pi$ [2]. Strong confirmation was presented by the GAMS group at the recent
LEAP'96 conference [7]. GAMS claim to observe a very strong $\eta (1295)$, 
as well as a smaller $\iota (1440)$ in $\eta \pi ^+ \pi ^-$; 
they express confidence that
the $\eta (1295)$ is distinct from $f_1(1285)$, which they also observe. Let us
assume $\eta (1295)$ to be the radial excitation of $\eta (550)$, and the 
partner of
$\pi (1300)$. A $0^-$ $K^*(1460)$ is reported by the Particle Data group,
though needing confirmation. If we use for guidance masses of the nearby $2^+$
nonet, containing $f_2(1270)$, $a_2(1320)$, $K^*(1430)$ and $f_2'(1525)$, the 
mass of $K^*(1460)$ fits well into a nonet with $\eta (1295)$ and $\pi (1300)$,
but one then expects the highest state of the nonet around 1550 MeV. 
(We shall refer to this as the $s\bar s$ state for
brevity, though it may be mixed with $q\bar q$.) We propose that this
state is pushed down in mass to 1430 MeV by mixing with the broad $0^-$
glueball, and the mixing causes both states to appear strongly in $J/\Psi$
radiative decays.

We now need to discuss mixing between the glueball and the $q\bar q$ nonet. 
The Mark III group shows that any $\eta (1295)$ signal in $J/\Psi \to \gamma
(\eta \pi \pi )$ is less than 20\% of the $\iota (1440)$ [8].
This implies that its mixing with a $0^-$ glueball component, which we denote
by $|G>$, is small. 
Since $|G>$ is an SU(3) singlet, it implies that $\eta (1295)$ is close to an
SU(3) octet.
We adopt a mixing scheme where $\eta (1295)$,
$\eta (1440)$ and the broad resonance are described respectively by states 
$|A>$, $|B>$ and $|C>$ where
\begin {eqnarray}
|A> &=& \cos \theta |q\bar q> + \sin \theta |s\bar s>,  \\
|B> &=& -\sin \theta \cos \phi |q\bar q> + \cos \theta \cos \phi |s\bar s> +
\sin \phi |G>,  \\
|C> &=& \sin \theta \sin \phi |q\bar q> - \cos \theta \sin \phi |s\bar s> +
\cos \phi |G>.
\end {eqnarray}

Next we wish to parametrize channel widths $\Gamma _i(s)$ of equn. (1). 
We describe decays to $\eta \pi \pi$ as $\eta \sigma$, where $\sigma$
is a shorthand for the whole $\pi \pi $ S-wave amplitude [9]; $\Gamma (s)$ is
evaluated by numerical integration over the available phase space. 
If instead we substitute decays to $a_0\pi$ or some combination of this 
with $\eta \sigma$, results change very little. 
The $a_0\pi$ and $\eta \sigma$ decays can be
separated only from details of the Dalitz plots, with which we are not
presently concerned. 
We include $|B> \to \eta \sigma$ and $\bar K(K\pi )_S$, where
$(K\pi )_S$ stands for the $K\pi$ S-wave; 
decays to $\rho \rho$ seem to be absent. 
Again, decays to $\bar KK^*$ give results similar to $\bar K(K\pi )_S$ and
within uncertainties about form factors. 
Finally we allow for decays of the
broad state $|C>$ to $\eta \sigma$, $\bar K(K\pi )_S$, $\rho \rho$, 
$\omega \omega$, $K^*\bar K^*$ and $\phi \phi$, although we in fact 
find the contribution from this broad
state to $\bar K(K\pi )_S$ or $K\bar K^*$ is compatible with zero. 

For resonance decays we include a form factor $B_i(k) = exp(-\alpha
k_i^2)$, where $k_i$ are momenta of decay products in the rest frame of the
resonance and $\alpha$ optimizes at 1 GeV$^{-2}$. 
For P-state production of $\rho \rho$, $\omega \omega$, $K^*\bar K^*$ and $\phi
\phi$, we use a phase space factor $k_i^3/(k_i^2 + \beta)$ where $\beta = 0.06$
(GeV/c)$^2$; this allows for a centrifugal barrier of radius 0.8 fm.
We also include a form factor $B_J(s) = \exp (-\alpha
q^2)q^3/(q^2 + \beta)$ for the decays $J/\Psi \to \gamma X$; 
here $q$ is the momentum
of resonance $X = A,B,C$ in the $J/\Psi$ rest frame. Our equations are thus
identical to those spelled out in detail by 
Achasov and Shestakov [10], except for inclusion of form
factors and the centrifugal barrier.
The $\Gamma _i(s)$ for each  decay are illustrated 
in Fig. 2. 
In calculating $\rho
\rho$ phase space, we find that interferences between different charge states,
e.g. between $\rho \to \pi ^+_1 \pi ^-_2,$ $\rho \to \pi ^+_3 \pi ^-_4,$ and 
$\rho \to \pi ^+_1 \pi ^-_4,$ $\rho \to \pi ^+_3 \pi ^-_2,$ have little effect
on $\Gamma (s)$. The same is true for $K\bar K^*$, $K^*\bar K^*$ and $\bar
K(K\pi )_S$. 

We fit data on $J/\Psi \to \gamma (\eta \pi \pi )$ and $\gamma (K\bar K \pi)$ 
from Mark III [4,8] and DM2 [11]. 
There are considerable uncertainties about absolute branching ratios. 
DM2 quote a branching ratio of the iota to $K\bar K \pi$ of $3.8 \times 10
^{-3}$ [11]. Mark III [4] find $
2.56 \times 10^{-3}$, but attribute some of it 
to $1^+$.  We correct DM2 results for this component and average with the Mark
III branching ratio to obtain 
$B[J/\Psi \to \gamma \iota (1440)].B[\iota \to K\bar K \pi] = 
2.1 \times 10^{-3}$. For $\eta \pi \pi$, we average the DM2 [11] and Mark III 
[8] results (before $a_0(980)$ cuts are applied) to obtain
$B[J/\Psi \to \gamma \iota (1440)].B[\iota \to \eta \pi \pi] = 0.55 
\times 10^{-3}$.
These values influence the mixing angle $\phi$, 
but have no effect on the quality of fits to data.
Secondly we fit Mark III data on $J/\Psi \to \gamma (\rho \rho )$ [5], 
$\gamma (\omega \omega )$ [12], $\gamma (K^*\bar K^*)$ [13] and 
$\gamma (\phi \phi )$ [13].  

The mass of the broad state in not well determined, but lies in the range 
1750--2100 MeV. 
There is some sensitivity to the form factor parameter
$\alpha$, mainly because of its effect through the $J/\Psi$ form factor 
$B_J(s)$; $\alpha$ must lie between 0.5 and 1.5 (GeV/c)$^{-2}$. There is
negligible dependence on the centrifugal barrier factor $\beta$.
Our earlier conclusions [1] on $0^+$ peaks at 1500, 1770 and 2100 MeV in 
$4\pi$ are completely unaffected by the re-parametrisation of the $0^-$
amplitude.

Fits to representative data are shown in Fig. 3. 
We assume flavour blindness, i.e. coupling
constants for $\rho \rho$, $\omega \omega$, $K^*\bar K^*$ and $\phi \phi$ in the
ratios 3:1:4:1. The first three channels agree well with this assumption and we
take this as evidence of the glueball character of the broad state. The 
glueball argument is strengthened by the observation that the broad state
couples to $\eta\sigma$ strongly.
There is evidence for the broad $0^-$ component in $\eta \sigma$ from Crystal
Barrel data on $\bar pp \to \eta \pi ^0\pi ^0\pi ^0$ [14].
In Fig. 3(a), we fit the low mass end of the $\eta \pi \pi$ spectrum, where
$0^-$ is likely to dominate; the fit fails at higher masses, where it is likely
that $1^+$ and/or $2^-$ will contribute.

For $K^*\bar K^*$, the data lie slightly above our fit near 1920 MeV. This
might be due to mixing with the $3^1S_0~s\bar s$ state expected near this mass,
or it could be due to $f_2(1920)$ which we discuss below. A spin-parity analysis
of the data is needed.

For $\phi \phi$,
there is a factor 4 discrepancy in branching ratio between DM2 [15] and Mark
III [13]. Our fit lies well above DM2 results, but agrees fairly well with
Mark III, Fig. 3(f). We give reasons below to believe that the peak at 2200
MeV has $J^P = 2^+$, in agreement with recent JETSET results [16].

Wermes [17] has attempted fits similar to ours, 
but using a resonance mass close to $\iota (1440)$.
He did not include the broad component in $\eta \pi \pi$ or $K\bar K \pi$. 
We find that the $\eta \pi \pi$  channel plays a significant role in 
getting the right
$s$ dependence for the $\rho \rho$ channel and good fits to $\omega \omega$ and
$K^*\bar K^*$.

In a recent preprint, Close, Farrar and Li [18] have derived a formula relating
the branching ratio for production of glueballs in radiative $J/\Psi $ decays 
to their widths (and masses). 
From Fig. 2, the mean total width of the broad $0^-$ signal is about 1.4 GeV,
when folded with the resonance shape. 
Using their formula, one predicts a branching ratio $J/\Psi \to \gamma 0^- = 
28 \times 10^{-3}$. 
We now compare that with our fit. For $K^*\bar K^*$ and $\omega \omega$
decays, Mark III quote integrated branching ratios of $5.6 \times 10^{-3}$ and
$1.7 \times 10^{-3}$. For $\rho \rho$, the measured branching ratio is $6.3
\times 10^{-3}$ integrated up to 2.35 GeV [1], and we fit an additional $2.0
\times 10^{-3}$ at higher masses. For $\phi \phi$ our fit gives $0.8 \times
10^{-3}$. The largest uncertainty is in the $\eta \pi \pi$ channel. Our fit
gives $5.6 \times 10 ^{-3}$ up to 2.0 GeV and a further $3.5 \times 10^{-3}$
above that. However, there are no spin-parity analyses to check these large
contributions from the broad signal in this channel. These values add up to
$25 \times 10^{-3}$, in agreement with prediction. However, there must be 
an uncertainty of at least 25\% in present branching ratios.

To determine the mixing angle $\phi$, we assume that only state $|G>$ is
produced in $J/\Psi$ decays. From the total branching ratio for the broad
state, $25 \times 10^{-3}$, and that of the narrow $\iota$, $2.65 \times
10^{-3}$, we find $\tan ^2 \phi = 0.106$, or $\phi = 18^{\circ}$. 
There is probably an error of $\pm 25\%$ in $\tan ^2 \phi$ from the branching 
ratios. We are unable to form
any accurate estimate of the mixing angle $\theta$ within the nonet, but the
Mark III data [8] suggest that $\iota (1440)$ is close to an SU(3) singlet.

If we adopt a value of 2.1 GeV for the glueball mass, together with our value of
$\tan \phi$, repulsion through the real parts of the Hamiltonian leads to an
unperturbed mass for the $s\bar s$ state of 1520 MeV.
This result is reasonable, but cannot be considered accurate because of
coupling through decay channels.
There is a strong dispersive effect on the rapidly rising edge of 
$|C> \to \rho \rho$ and $\eta \sigma$. 
It seems quite natural that a balance between this attraction and the repulsion
between levels will result in a mass for state $|B>$ on the rising edge of the
$\rho \rho$ threshold, as observed experimentally. 

The Obelix group has reported evidence for another $0^-$ resonance at 1460 MeV
[19]. There have likewise been suggestions for a $K\bar K^*$ 
resonance at 1460 MeV by DM2 [20] or at 1490 MeV by Mark III [4]. 
A second narrow state at this mass does not fit naturally into our scheme.
However, we find that the apparent mass of the narrow resonance can  be 
shifted appreciably because of differing amplitudes and phases of the 
broad and narrow components in different reactions. 
Our belief is that a conclusive analysis of Dalitz plots is only possible if
the experimental groups will pool their data and re-analyse all channels with a
single parametrisation including the $s$-dependence of channel widths. This
analysis needs to include the broadening of the $K^*$ proposed by Frank et al.
[21]. Such a re-analysis is now needed.

We conclude with some conjectures about the $2^+$ glueball, based on analogy
with 
our observations for $0^-$. The $2^+$ glueball is 
predicted by
Lattice QCD calculations to be a factor 1.5 heavier than the $0^+$ glueball and
to lie around 2200 MeV. It is likely to mix with $^3F_2$ and $^3P_2~q\bar q$
and $s\bar s$ mesons. We shall show that the available data on $2^+$ states
are consistent with a broad $2^+$ glueball centred roughly at 2000 MeV and with
a width of $\sim 650$ MeV. The central ideas are that (i) the glueball pushes
$f_2(1565)$ and $\theta (1710)$ to masses lower than expected for the $2^3P_2$
nonet; (ii) the $\phi \phi$ states observed by Etkin et al. [22] 
are consistent with
mixing on the upper edge of the glueball; and (iii) the observations of $2^+$
signals at 1920--1950 MeV are consistent with a $3^3P_2$ state near the peak
of the glueball and mixing with it.

Clues to this scenario come from expected masses for radial excitations. 
If a Regge spacing of 1.1 GeV$^2$ in $s$ is assumed, the first
and second radial excitations of $f_2(1270)$ are expected at 1650 and 1950 MeV,
and the radial excitations of $f_2'(1525)$ at 1850 and 2125 MeV. The first of
these numbers is supported by new 
evidence for an $I = 1$ $J^P = 2^+$ state in $\eta \pi$ at 1650 MeV [23].
There is now a well documented $2^+$ $I = 0$ resonance formerly known as $AX$ 
at 1540--1555 MeV, coupling to $\pi \pi$, $\rho \rho$ and 
$\omega \omega$ [2,24]. We interpret this
as the radial excitation of $f_2(1270)$. 
The low mass of this state might arise from attraction to the sharp 
$\omega \omega$ threshold.
However, another possibility is that the $2^+$ glueball may push the 
radial excitation down to the $\omega \omega$ threshold. If so, a similar or
larger effect should be observed for the radial excitation of $f_2'(1525)$.
It is tempting to identify this with $\theta (1710)$, though it is
presently disputed whether this resonance has $J^P = 0^+$ or $2^+$. 
The data which are statistically strongest [25] favour $2^+$. The strong
appearance of this resonance in $J/\Psi \to \gamma (K\bar K)$ would be a natural
parallel to the appearance of $\iota (1440)$ in $J/\Psi $ radiative decays.
We propose that the masses of both states lie on the sharply rising edge of 
broad glueballs. 

At higher mass, there is good evidence from both GAMS [26] and VES [27] for
$f_2(1920)$ decaying to $\omega \omega$, and from the Omega group in $4\pi$
[28]. There is also evidence from LASS [29] for
$f_2(1950)$ decaying to $K^*(892)\bar K^*(892)$. We suggest that these are all 
the same resonance, and the presence of both strange and non-strange 
decay modes is evidence for the second
radial excitation of $f_2(1270)$ mixing with the $2^+$ glueball. 

Mixing of a broad glueball with further $^3P_2$ $q\bar q$ radial excitations
and $^3F_2$ $q\bar q$ states could explain the anomalously strong signals
observed in $\phi \phi$ by the JETSET collaboration [16] and by 
Etkin et al. [22], who claim to observe three resonances.
JETSET observe a peak centred at 2180-2200 MeV. There is further evidence
for a $2^+$ state around 2200 MeV [2]. We note that the data of 
Etkin et al. show a peak at 2150 MeV, though the pole position is reported
as being at $2011 ^{+62}_{-76}$ MeV.  
Etkin et al. observe this peak in the $\phi
\phi $ S-wave. This seems to us a good candidate for
glueball mixing with the $3^3P_2$ $s\bar s$ state.

Etkin et al. also report a narrow $\phi \phi ~2^+$ D-wave resonance at 2297 MeV.
This is close to the right mass for the $^3F_2$ ground state $s\bar s$ state.
The D-state resonance is produced much more weakly than the S-state, and we
comment on a possible reason. Lattice QCD calculations suggest that the
glueballs have radii distinctly smaller ($\sim 0.3 $ fm) 
than $q\bar q$ states [30]. The high masses of the glueballs compared to
$q\bar q$ ground states also point towards high constituent gluon mass, 
hence small radii. 
The $3^3P_2$ state, with wave function $\propto r$ for small $r$, 
will overlap much better with a small object than will $1^3F_2$, which has a
wave function $\propto r^3$. 

Etkin et al. also claim a broad D-wave 
resonance. We suggest that this is a
parametrisation of the broad $2^+$ contribution from the glueball. Their data
fall rapidly above 2350 MeV, leading us to believe that the glueball falls
rapidly there. In summary, we propose that $\theta (1710)$ flags the steeply
rising edge of the glueball and the $\phi \phi$ data exhibit its falling edge.
The result is a central mass of $\sim 2000$ MeV and a width of order 650 MeV.

Our scheme identifies all $I = 0 ~q\bar q~ 2^+$ states in this mass range
except $1^3F_2~ q\bar q$.
It is unlikely to mix much with a compact glueball, because of its F-state
$q\bar q$ wave function and perhaps because of octet character.

For a $2^+$ glueball 650 MeV wide, Close, Farrar and Li predict a branching
ratio of $33 \times 10^{-3}$ for $J/\Psi $ radiative decays. Where is this huge
signal hiding? Our analysis of $J/\Psi \to \gamma (4\pi )$ locates only $0.8
\times 10^{-3}$ for production of $f_2(1565)$ or $f_2(1710)$ [1] and $\Theta
(1710)$ contributes only $0.97 \times 10 ^{-3}$ through its $K\bar K$ decay
[2]. The obvious place to look for the missing signal is in the channels $\eta
\eta$, $\eta \eta '$ and $\eta '\eta '$, which have not been studied
extensively.

We are grateful to Dr. Frank Close for advice and discussion on branching
ratios.

\newpage
\begin{figure}[htbp]
\begin{center}\hspace*{-0.cm}
\epsfysize=15.0cm
\epsffile{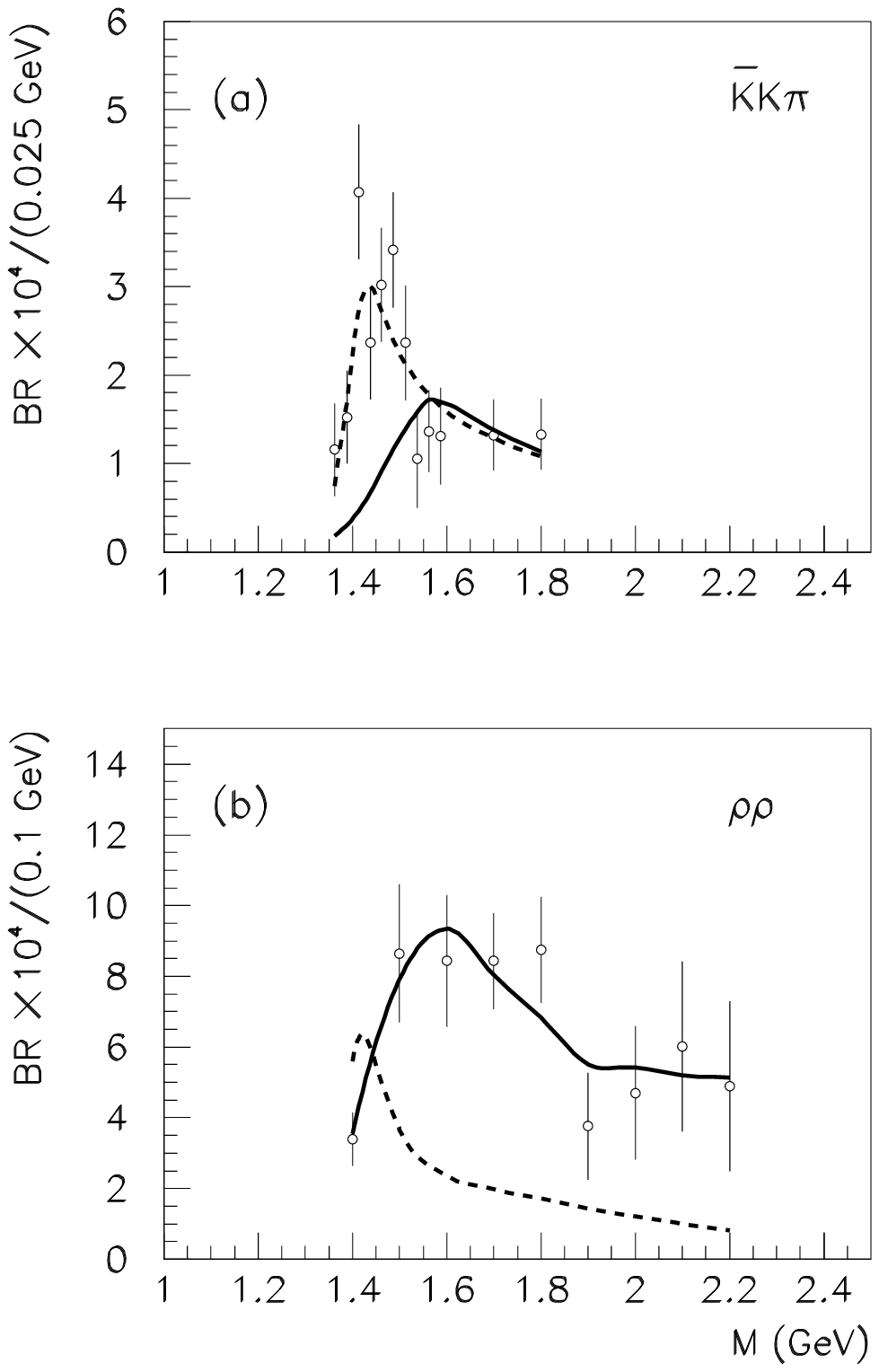}
\end{center}
\caption{A fit to Mark III data on (a) $J/\Psi \to \gamma (K\bar K\pi)$ and 
(b) $J/\Psi \to \gamma (4\pi )$ with a single resonance. The dashed curve shows
the best fit to (a) and the full curves the best fit to (b).
}
\label{fig1}
\end{figure}

\newpage
\begin{figure}[htbp]
\begin{center}\hspace*{-0.cm}
\epsfysize=15.0cm
\epsffile{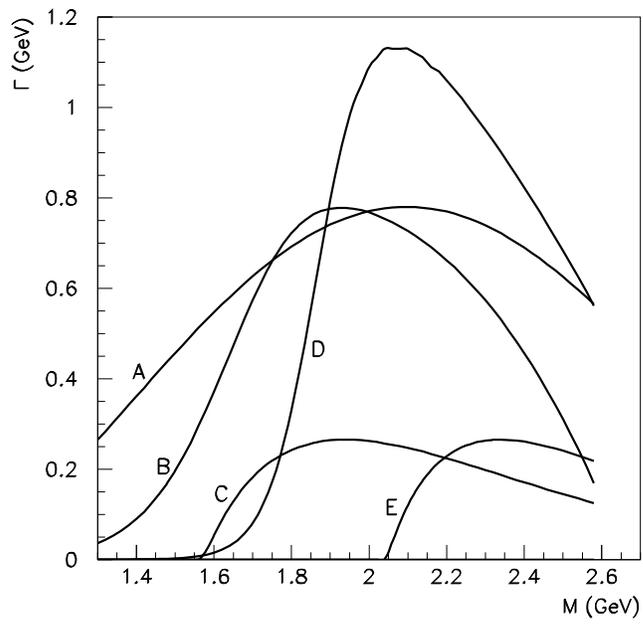}
\end{center}
\caption{$\Gamma (s)$ with $\alpha = 1$ GeV$^{-2}$, and  normalizations fitted
to the data. Curves show $\eta \sigma$ (A), $\rho \rho$ (B),
$\omega \omega$ (C), $K^*K^*$ (D), $\phi \phi$ (E).
}
\label{fig2}
\end{figure}

\newpage
\begin{figure}[htbp]
\begin{center}\hspace*{-0.cm}
\epsfysize=15.0cm
\epsffile{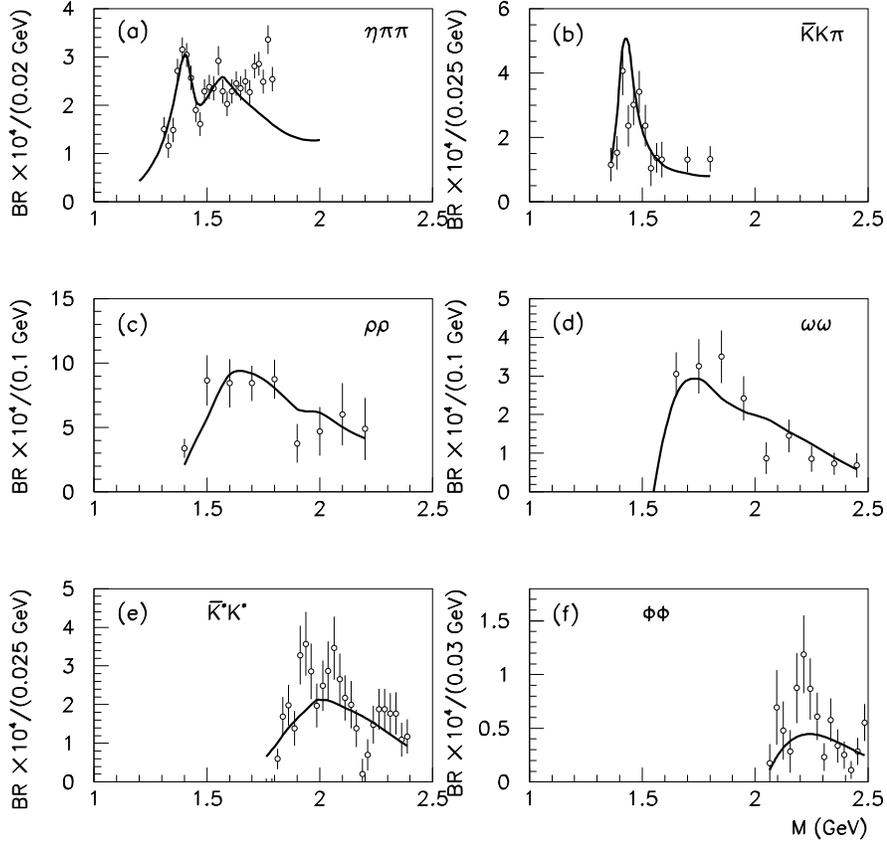}
\end{center}
\caption{Fits to data on radiative $J/\Psi $ decays, taking $M = 1800$
MeV for the broad state $|C>$: (a) $\eta \pi \pi$ from DM2, 
(b) $K\bar K \pi$, (c) $\rho \rho$, (d) $\omega \omega$,
(e) $K^*K^*$, and (f) $\phi \phi$. Data for (b)--(f) are from Mark III.
}
\label{fig3}
\end{figure}

\end {document}